\documentstyle[sprocl]{article}
\input{psfig}

\def\Journal#1#2#3#4{{#1} {\bf #2}, #3 (#4)}

\def\PRL{\em Phys. Rev. Lett.}
\def\PRB{{\em Phys. Rev.} B}

\def\be{\begin{equation}}
\def\ee{\end{equation}}
\def\bea{\begin{eqnarray}}
\def\eea{\end{eqnarray}}

\begin{document}
\def \rem #1 {{}}
\newcommand{\CC} {{\cal C}}
\newcommand{\DD} {{\cal D}}
\newcommand {\HH} {{\cal H}}
\newcommand {\x} {{\bf x}}

\title{CLUSTER MAXIMIZATION, NON-LOCALITY, AND RANDOM TILINGS}

\author{ C.L. HENLEY}

\address{Dept. of Physics, Cornell University,\\
Ithaca, NY 14853-2501, USA}

\maketitle\abstracts{
Jeong and Steinhardt (JS) 
implement local rules by selecting the sub-ensemble of 
tilings which
have the maximum occurrences of a chosen pattern (``cluster'') ${\cal C}$.
It is unknown how to prove that a given $\cal C$
implies a given sub-ensemble; counterexamples given here 
demonstrate this problem is nonlocal, and show the JS results
depend on periodic boundary conditions. 
Sub-ensembles which are 
{\it random} tilings of supertiles
are at least as interesting, mathematically and physically, 
as the ideal cases emphasized by JS; 
the case $\CC =$ star-decagon is explored here. 
Finally,  I suggest {\it minimizing} the frequency of chosen intersite distances
as a more physical alternative to the cluster approach. 
}

\section{Introduction}

Although the ``random-tiling'' and ``local-rules'' scenarios
give rather different pictures of the stabilization of quasicrystal order,
they are hard to distinguish since they both predict Bragg peaks
and quasiperiodicity in projections of 3D slabs.~\cite{hen-rtart}
Before this question can be settled, we must understand
how the microscopic structural energy favors the quasicrystal structure. 
One common assumes a decorated tiling, implicitly 
assuming the tilings are all degenerate (= random tiling) 
within a  rather coarse energy resolution, as is consistent
with existing pseudobinary pair potentials.~\cite{mvic} 
On a finer energy scale, the tile-tile effective Hamiltonian
might either favor a periodic crystal (so the quasicrystal is
an entropically stabilized random tiling), 
or else a quasiperiodic tiling.
Some interatomic potentials are known~\cite{FK} that (in simulations)
produce decorated quasicrystal random tilings upon cooling. 

Jeong and Steinhardt~\cite{JS-PRL}
(JS) idealized the tile Hamiltonian so as to define
restricted tilings:
in their simplest cases, 
they choose a ``cluster'' $\CC$ and take the subsensemble of 
tilings that maximize its frequency.
I will call such a prescription a ``maximizing rule'' (in analogy to
``matching rule''.)
The tiling is specified much more economically this way
than by enumerating the complete ``$R$-atlas'' of allowed
local patterns~\cite{levitov}.
For several different cases,  
JS inferred the subensemble from images of the ground states
obtained in simulations. 
The fact that JS obtained random tilings
while searching for quasiperiodic ones suggests that the former
cases are more common than the latter. 

The original JS paper~\cite{JS-PRL} -- and work it inspired~\cite{gahler} -- 
has a major gap: 
we don't know how to prove mathematically that the subsensemble is
optimal for the given Hamiltonian.  
JS in fact accomplished such a proof~\cite{JS-PRB}, 
but only for a certain (fairly large) $\CC$, copies of which can {\it cover}
the entire tiling (with overlaps allowed).~\cite {gummelt}
This gap is shared with most {\em atomic} quasicrystal models;
a {\em general} understanding of the relation of Hamiltonian to tiling 
would have immediate applications 
in predicting quasicrystal structures.\cite{mvic,FK}

Among the examples found by JS, 
\cite{JS-PRL}$^{\!,\,}$\cite{JS-PRB}
symmetric cluster rules such as $\CC_{B/T}$ or the star-decagon $\CC_{SD}$
gave random tilings;
quasiperiodicity could be forced only for rather special
$\CC$'s that break the 5-fold rotational symmetry. 
Conceivably this could be realized 
by vacancy alternations in the inner $\rm Al_{12}$ shell of a
Mackay Icosahedron, 
or in the ``$\delta$'' (six-dimensional body center) sites~\cite{mvic}.

\begin{figure}
    \begin{minipage}[b]{.44\linewidth}
    \psfig{figure=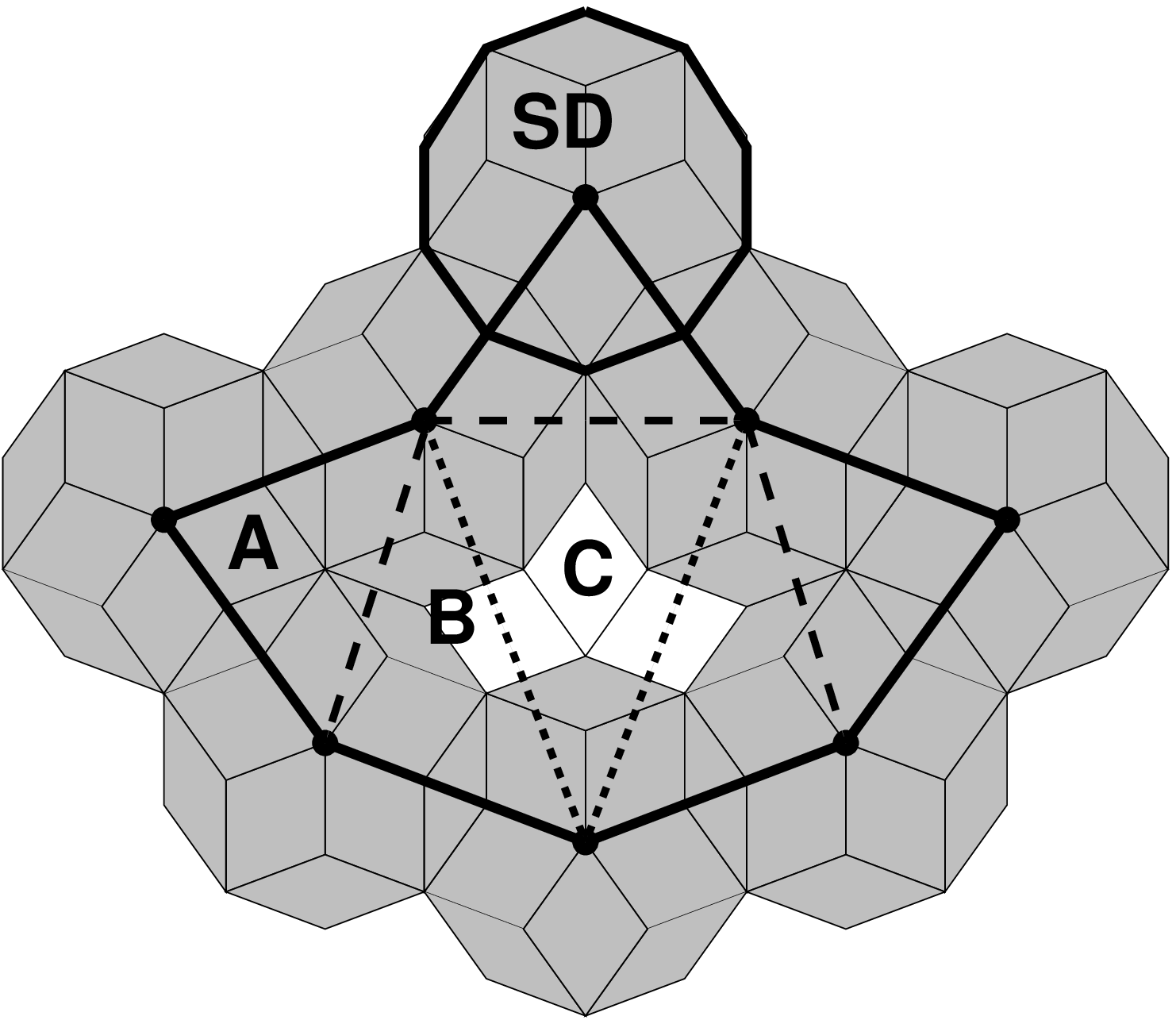,height=1.6in,width=\linewidth}
\end{minipage} \hfill
\begin{minipage}[b]{.44\linewidth}
    \centering\psfig{figure=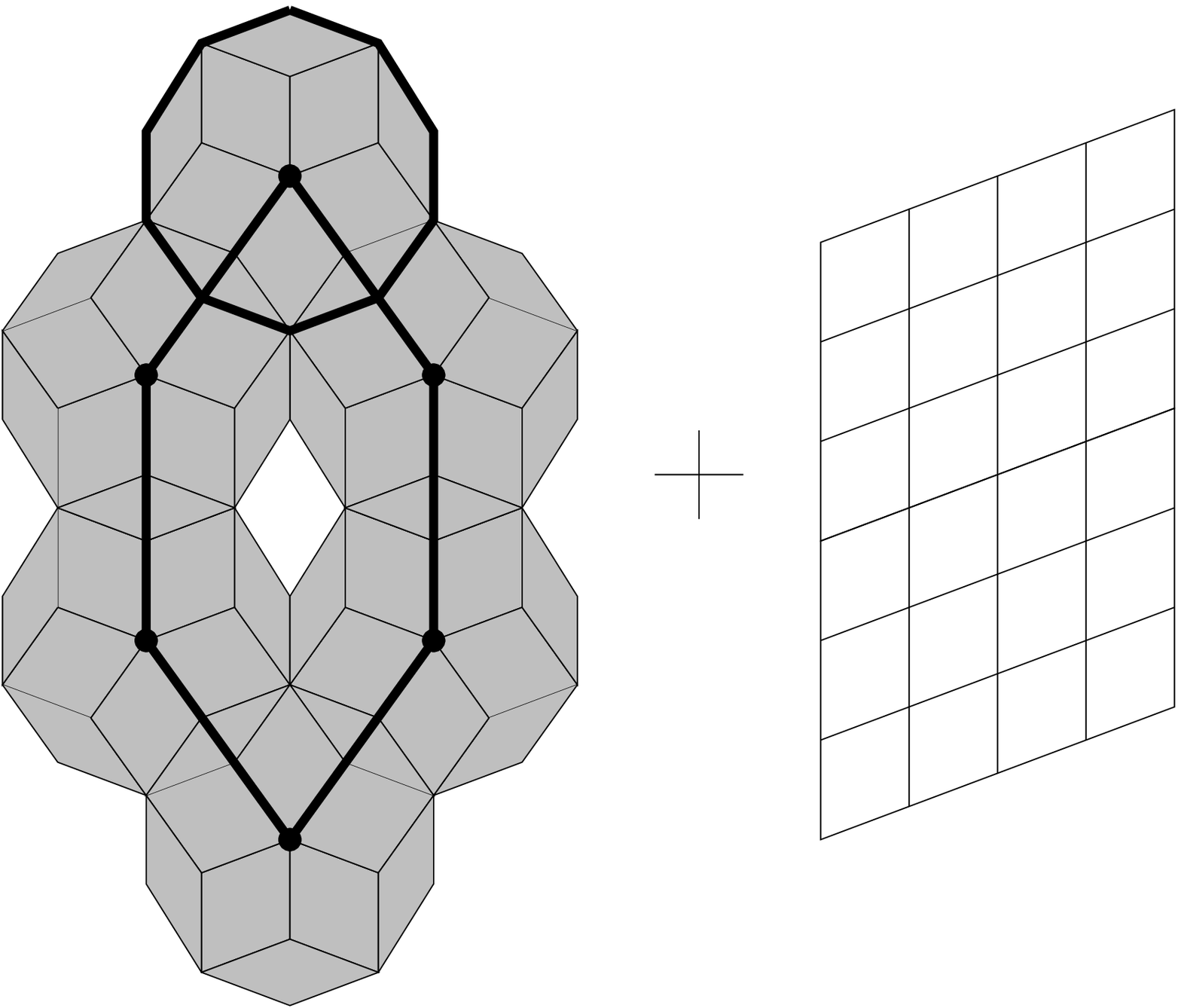,height=1.6in,width=\linewidth}
\end{minipage}
\caption{(a). Tiles of SD packing (centers marked; one SD is outlined), 
(b).  Unit cell of packing with maximum cluster density (hexagon tile), 
coexisting with domain of fat rhombi.}
\label{fig-SD}
\end{figure}

\section{Supertiling maximizing the SD-cluster}

The random tiling cases are, perhaps, even harder than the
quasiperiodic ones, since the optimum is not a quasi-unique
pattern, but a degenerate ensemble.
Such cases might model changes in symmetry due to supertile formation
which give superlattice-like diffraction.\cite{lancon}
As an explicit example, I now specialize to the ``star-decagon'' cluster
$\CC_{SD}$
(called ``$\CC_{D}$'' by JS~\cite{JS-PRL}).


The random tiling which describes the JS results for $\CC_{SD}$
is formed by the supertiles A,B,C outlined in 
Fig.~\ref{fig-SD}(a), which add to form a ``boat'' shape. 
Supertiles 2A+2B can be combined to make a ``hexagon'' 
as in Fig.~\ref{fig-SD}(b).
(Here we can draw the B triangle
hypotenuses in two ways in the interior, 
but we must count them both as the same tiling.)
One could also create a 5-pointed ``star'' with $\CC_{SD}$ on each vertex
(the top part of Fig.~\ref{fig-SD}(a) is a fragment of this star.)
The A-B-C tilings have as a subset
the random tilings of boat, hexagon, and star 
(which I called ``two-level tiling''~\cite{hen-rtart}), which in
turn have include the Penrose single-arrow-edge tiling.
Using this, the ideal SD frequency in a connected tiling is found
to be $\tau^{-3}/\sqrt 5$ clusters per vertex. 

However, these tilings are 
optimal only under the constraint that the tiling is {\it connected}, 
{\it i.e.}  every tile edge is fitted against another tile edge 
with no gaps.
Let us try instead a more physically relevant game. 
Given many fat and skinny rhombi, with the ideal number ratio 
$\tau\equiv (1+\sqrt{5})/2$, what packing has the most clusters?
It is actually a {\it phase-separated} configuration, consisting
of a domain of hexagons, plus  a domain of fat rhombi 
(Fig.~\ref{fig-SD}(b)).~\footnote
{Such an instability appears less likely when the energy has a cusp
as a function of phason strain 
as observed by JS~\cite{JS-PRL} 
in the quasiperiodic (non-random tiling) cases
(nonzero value of their ``$\alpha_2$''~\cite{JS-PRL}) 
-- but even then it cannot be ruled out.}
For every tile, we have $1/7\tau^2$ hexagons (each with two SD clusters), 
and $\sqrt 5/7 \tau^6 =0.018$ Fat tiles left over, giving a frequency
of $2/7\tau^2$ clusters per tile which is 3.3\% better than the 
connected tiling.

For a covering cluster, an essentially local proof of optimization 
is workable~\cite{JS-PRB} 
(extended by a clever use of inflation). 
Contrariwise, in the $\CC_{SD}$ case 
any {\em local} proof, 
based on counting and dividing areas, can be ruled out:
the ``phase separated'' state of Fig.~\ref{fig-SD}(b) always
beats the random tiling. 
The optimization is inherently nonlocal; a proof must
make essential use of the
connectedness of the tiling, or the requirement that the mean phason
strain be zero. 

As an aside, 
I have studied the tilings defined by a covering rule:
{\it every skinny rhombus must be covered by (at least one) SD}. 
The A-B-C random tilings have this property. 
It is easy to show that, if there is even one adjoining pair of 
{\em parallel} Fat tiles, 
then there is at most one SD in the whole tiling.
Excluding that, it is then easy (by enumerating the possible surroundings 
of a corner of the SD cluster) to show we obtain
just the random tiling made of the tiles in Fig.~\ref{fig-SD}(a) 
together with one added kind of tile, a large star.




\section{Critique of ``cluster'' picture and conclusion}

Do max-rules really have a more natural 
connection to an  interatomic Hamiltonian than Penrose matching rules do?
The total energy of a structure defines an effective potential of the
atomic coordinates which -- we assume -- is dominated by
few-body (up to 4-body) interactions among near neighbors, 
and two-body interactions to further neighbors. 
To favor a $\CC$-cluster of $M$ tiles (and {\it not}\  favor
smaller fragments of that cluster) requires a potential
term among atoms on $O(M)$ tiles, that is among {\em at least}
$O(M)$ atoms~\cite{hen-rtart}.
That seems implausible, although it is just what is postulated in 
models with a hierarchical 
electronic structure.\cite{janot} 
 

I suggest instead to explore a version of the JS game which
I'll call ``constraint'' rules. 
Rather than favor a chosen local pattern, let us {\it disfavor} 
certain chosen inter-vertex distances and take the sub-ensemble
in which they are forbidden (or if that is too restrictive, 
in which their frequency is minimized.)
A few such constraints
can very effectively limit the allowed local patterns~\cite{growth}. 
So, just like the ``cluster'' rules, this 
offers a more economical specification than the $R$-atlas.~\cite{levitov} 
Again one expects that random super-tilings are more generic than
quasiperiodic ones.
And constraint rules are much more plausible, a priori, since they
are naturally derived from interatomic {\it pair} Hamiltonians. 



\section*{Acknowledgments}
This work was supported by DOE contract DE-FG02-89ER-45405.
I thank H.-C.~Jeong and P.~J.~Steinhardt for discussions and
for sending unpublished images from their simulations.

\end{document}